# Systematic Benchmarking of Macrosegregation: The Performance of a Modified Hybrid Model

Ali Moeinirad and Ehsan Amani[*]

*Department of Mechanical Engineering, Amirkabir University of Technology (Tehran Polytechnic), Iran*

**Abstract**

Recently, a new alloy solidification benchmark, called AFRODITE, with well-defined setups and state-of-the-art measurements has emerged, enabling a thorough assessment of MacroSegregation (MS) solvers, particularly in terms of their ability to predict different features of MS maps. In this research, we first develop an analytical solution for the alloy-solidification Stefan problem, which involves melt, solid, and mushy regions. This new analytical solution extends a previous solution (S. Cho and J. Sunderland, "Heat-conduction problems with melting or freezing", J. Heat Transfer, vol. 91, pp. 421-426, 1969) by incorporating a linear microsegregation law as a function of temperature in place of spatial coordinate. Then, we adopt this solution to verify an OpenFOAM MS solver in a limiting condition, where only heat diffusion is present. Subsequently, to capture the MS map of the Sn-3%Pb AFRODITE benchmark, the solver is incorporated using the standard Blake-Kozeny-Carman permeability law and one of its hybrid variants, slightly modified in this work to better align with physics by ensuring a continuous transition of characteristics from the slurry to the porous regions of the mush. It is demonstrated that the hybrid model predicts the main features of the MS map, including the channel segregates morphology and peak segregation degree due to the pile-up effect, in much finer agreement with the experimental observation. Careful analyses of the results reveal that these improved predictions stem from the hybrid model's more accurate estimation of the re-melting, melt flow advection parallel, and advection normal to the solidification front.

---
[*] Corresponding author. Address: Mechanical Engineering Dept., Amirkabir University of Technology (Tehran Polytechnic), 424 Hafez Avenue, Tehran, P.O.Box: 15875-4413, Iran. Tel: +9821 64543404. Email: eamani@aut.ac.ir

# 1. Introduction

Non-uniformity in the solute spatial distribution, also known as MacroSegregation (MS), is one of the major defects in alloy casting. This phenomenon can be attributed to the relative motion of the solid and liquid phases or fluid flow in the mushy zone which originates from miscellaneous sources such as natural or forced convection, shrinkage in the liquid phase, grain movement, and deformation of the solid skeleton [1]. For a review of the physical mechanisms of MS in different kinds of casting process, readers can consult Ref. [2]. MS reduces the product quality by degrading its homogeneity and mechanical properties. Since MS cannot be eliminated entirely during heat treatment, preventing or at least reducing it in the casting process is crucial [3]. Channel segregation or freckling, characterized as long, narrow, highly solute-rich trails oriented in a preferred direction, represents a severe defect in the final cast product, which is difficult to eliminate through any secondary manufacturing process [4, 5]. Due to the harsh casting environment, expensive measurements, and the time required for each test, plant-scale trials are seldom practical, and the computational modeling for the study of alloy casting becomes extremely valuable.

Through a pioneering study, Flemings and Nereo [6] developed an analytical model, called the local solute redistribution equation, to study the MS caused by inter-dendritic flow. The next generation of models, accounting for mass, momentum, and heat transport, was proposed based on either a single-fluid [7], also known as Enthalpy-Porosity (EP), or volume-averaged two-phase melt-solid [8] models. To account for Columnar-to-Equiaxed Transition (CET), Wu and Ludwig [9] derived a three-phase melt/columnar-solid/equiaxed-solid model and then extended it to a five-phase model [10, 11] accounting for two additional phases, namely the inter-dendritic melts confined in the columnar and equiaxed crystals. On the other hand, Wang, et al. [12] considered the melt, inter-dendritic liquid, and solid in their three-phase model. Increasing the number of phases escalates the computational cost while not necessarily improving accuracy, due to the increased number of approximate closures required for the interaction between phases.

Original EP or two-phase models were unable to account for the difference between the regions of porous-medium-like columnar dendrites and slurry-like freely moving equiaxed crystals within the mushy zone which significantly influences the convection inside this zone. This phenomenon was subsequently included through hybrid permeability models which switch between different laws based on some criteria. Chang and Stefanescu [13] developed an EP-based hybrid model that uses a pseudo-viscosity, increasing along the mushy zone and switching to Darcy's law at a critical solid mass fraction, called Dendritic Coherency Point (DCP). Ilegbusi and Mat [14] adopted a non-Newtonian viscosity in the slurry and Darcy's law in the columnar porous region based on DCP. Seredyński and Banaszek [15] pointed out that due to the complexity of the process and its dependence on miscellaneous factors, the validity of a constant threshold used in DCP for separating the columnar from equiaxed dendritic regions in the mushy zone is highly questionable. They proposed a hybrid EP/Front-Tracking (FT) strategy where the location of the surface between these regions is tracked by a virtual front propagating based on predefined dendrite tip kinetics. Multiscale approaches–replacing the macroscopic microsegregation closure with a mesoscale model like Cellular Automaton (CA) [16], Monte Carlo (MC) [17], or Phase Field (PF) [18]–are also considered tractable strategies for large-scale problems. They offer higher accuracy and levels of information, including grain structures, albeit at significantly much higher computational costs.

Regardless of the modeling procedure, MS models should be validated carefully against reliable benchmarks. Due to the complexity of MS phenomena, analytical solutions are scarce and limited to 1D overly simplified conditions [19-22]. Therefore, experimental databases providing MS maps are necessary for the assessment process.

There are two major experimental setups which, as a benchmark, substantially contributed to the study of alloy solidification. The first is the experiment by Hebditch and Hunt [23] on the solidification of Sn-Pb alloys in a rectangular mold. This database provided the MS map by the measurements of the MS degree at the end of solidification for two different alloys: Sn-5%Pb and Pb-48%Sn. Using an EP modeling approach, Kumar, et al. [24] assessed 3 variants of Blake-Kozeny-Carman (BKC) permeability laws against this database. Only one of these variants was able to correctly predict channel segregation. Chen and Shen [25] obtained a finite element solution using EP with the Lever rule and standard BKC law. Their predicted MS map showed better compatibility with the experimental measurements than the previous numerical computations using the same model [26]. The difference in these predictions was attributed to the distinction between the numerical algorithms, i.e., the fractional step method in Ref. [25] versus the penalty method in Ref. [26]. Seredyński and Banaszek [15] applied their hybrid EP-FT model to this benchmark and investigated the effect of permeability laws. They highlighted the sensitivity of the number and morphology of the predicted channel segregates to the permeability formula and the assumed size of the dendrite arm spacing. In addition, a higher tendency to channel formation was reported using anisotropic permeability laws. Kumar, et al. [27] carried out an EP-based simulation of this benchmark to explore the formation of channel segregates. They reported that a mesh resolution of at least two times the secondary arm spacing is required to resolve the flow in channels and predict their morphology and locations accurately. In addition, they compared the predictions by the Lever and Scheil microsegregation models, combined with the standard BKC permeability. They observed that the Scheil model predicts a higher segregation degree while the number of channels is the same using both models. They also demonstrated that the number of channels decreases by increasing the ratio of inertial to Darcy's drag in the modeling closure.

The need for newer experimental benchmarks on alloy solidification with more well-defined boundary conditions as well as MS measurements of lower uncertainties led to a series of experiments, based on the AFRODITE setup, initiated by Quillet, et al. [28]. Hachani, et al. [29] provided the MS map for Sn-3%Pb alloy using this setup. This map was then adopted by Carozzani, et al. [30] who compared an EP-based model using the Lever microsegregation rule with a multiscale EP-CA model, replacing the Lever rule with a CA to account for the description of the grain structures. They manifested that the recalescence in the temperature track was only captured by the multiscale model. However, the channel segregation predicted by both models was lower than that in the experiment. The probable cause of this discrepancy was attributed to the use of the isotropic BKC permeability in both models. Boussaa, et al. [31] adopted an EP-based model with the Lever rule and standard BKC closure for the same benchmark and reported that 3D simulations better capture the topology of the channels due to their 3D tubular configurations.

Hachani, et al. [32] extended the previous measurements on the AFRODITE setup to different alloy nominal concentrations of Sn -3%Pb, Sn -6.5%Pb, and Sn -10%Pb. The Sn-10%Pb test case was then used by Zheng, et al. [33] to evaluate an extended three-phase model [9] in predicting the as-cast structures. They pointed out that the extension to incorporate the crystal fragmentation of columnar dendrites is critical for accurate prediction of the Sn-10%Pb case study.

Hachani, et al. [34] and Khelfi, et al. [35] provided MS maps for the Sn-10%Pb alloy under the stirring effect of traveling magnetic fields. Wang, et al. [36] pointed out the need for a turbulence closure to simulate these cases. They used their previously developed three-phase model [12] in conjunction with a k-ε turbulence closure. However, the validity of interphase interaction closures under a turbulent flow condition as well as turbulence damping within the mushy region remained unanswered. Zhang, et al. [37] evaluated their extended three-phase model [33] using the same benchmark. They showed that the traveling magnetic field promoted the formation of equiaxed grains by the fragmentation process and facilitated the appearance of CET. The MS degree and channel formation increased by applying the traveling magnetic field for the chosen test case.

Despite the aforementioned progress, much more effort is still necessary to evaluate the performance of different modeling strategies to capture various features of MS maps of these standard benchmarks, especially for the brand new AFRODITE database which attracted much interest from the casting community and the previous computational studies on each of its test cases are limited. The contribution of the present research is two-fold: To perform a systematic validation and assessment of MS models, first, an extended analytical solution for a simplified alloy solidification problem is obtained and used to verify our solver in the limit case. This extended solution is based on a microsegregation law in terms of temperature, which is in accordance with widely-used laws in numerical models, rather than the spatial coordinate in the original solution. Second, the MS characteristics of the Sn-3%Pb AFRODITE test case are investigated and analyzed carefully, with the focus on the evaluation of the capability of the standard BKC permeability model and one of its hybrid variants in predicting positive and negative segregation and the number and morphology of the channel segregates. Note that we slightly modified the original hybrid model so as to incorporate a continuous variation of the relative advection terms from the slurry to porous regions.

The rest of the paper is organized as follows: Section 2 provides the governing equations comprising different sub-models of the present MS solver. Section 3 introduces the test case of the AFRODITE benchmark, used in this study. Section 4 outlines the numerical techniques for the solution of the governing equations. The systematic assessment of the present MS solver, including the verification against the new analytical solution, along with the analyses of MS characteristics using different modeling procedures are presented in section 5. Finally, the main conclusions of the work are summarized in section 6.

## 2. Mathematical modeling

### 2.1. Continuity and momentum

To study MS in an alloy solidification, the governing equations encompass fluid flow, heat transfer, species transport, and phase change phenomena. The flow equations involve variable-density continuity and momentum, which can be expressed as follows [38]:

$$\frac{\partial \rho}{\partial t} + \nabla \cdot (\rho \boldsymbol{u}) = 0, \tag{1}$$

$$\frac{\partial (\rho \boldsymbol{u})}{\partial t} + \nabla \cdot (\rho \boldsymbol{u}\boldsymbol{u}) = \nabla \cdot (\mu_{\text{eff}} \nabla \boldsymbol{u}) - \nabla p - \rho_0 \boldsymbol{g}[1 - \beta_T(T - T_0) - \beta_C(C_l - C_0)] + \boldsymbol{S}_u + \boldsymbol{S}_r, \tag{2}$$

where $\rho$ is the density, $p$ the pressure, $\boldsymbol{u}$ the velocity, and $\boldsymbol{g}$ the gravitational acceleration. The subscripts $l$ and $s$ refer to the liquid (melt) and solid states, respectively, and the average or mixture properties are indicated without subscripts. The bold symbols are used to indicate vectors and tensors. The mixture density can be computed by

$$\rho = g_s \rho_s + (1 - g_s)\rho_l, \tag{3}$$

where $g_s$ is the solid volume fraction (within the mushy region). The relation between solid volume fraction and solid mass fractions, $f_s$, is described by

$$\rho_s g_s = \rho f_s = (\rho_l g_l + \rho_s g_s) f_s = \left(\frac{f_s}{\rho_s} + \frac{f_l}{\rho_l}\right)^{-1} f_s, \tag{4}$$

In Eq. (2), $\mu_{\text{eff}}$ is the effective viscosity which equals the liquid molecular viscosity, $\mu_l$, assuming laminar flow. The Boussinesq approximation has been considered to represent thermal and solutal buoyancy; Therefore, $\beta_T$ and $\beta_C$ show the thermal and solutal expansion coefficients, $T$ and $C_l$ are the local temperature and solute concentration in the liquid phase, $C_0$ and $T_0$ indicate the corresponding reference values taken as the initial alloy concentration and initial temperature respectively, and the liquid and solid density in Eq. (3) are assumed constant. $S_u$ is the momentum source term that imposes the effect of the drag force in the mushy zone and can be expressed as follows:

$$S_u = -\frac{\mu_l}{K}\frac{\rho}{\rho_l}(\boldsymbol{u} - \boldsymbol{u}_s), \qquad (5)$$

where $K$ is the permeability of the mushy region, and $S_r$ is the relative advection given by, knowing $\boldsymbol{u} = f_s\boldsymbol{u}_s + (1 - f_s)\boldsymbol{u}_l$,

$$S_r = -\nabla \cdot \left[\rho \frac{f_s}{(1 - f_s)}(\boldsymbol{u} - \boldsymbol{u}_s)(\boldsymbol{u} - \boldsymbol{u}_s)\right]. \qquad (6)$$

To calculate the permeability, the growth morphology of the mushy region has to be taken into account. Here, two models are incorporated for this purpose.

(a) The standard BKC model

For alloys, the permeability as a function solid mass fraction is usually given by the Blake-Kozeny-Carman (BKC) equation [7, 39], which is based on the Darcy law in a porous medium,

$$K^{-1} = K_0^{-1}\left[\frac{f_{s,B}^2}{(1 - f_{s,B})^3}\right]; \quad K_0 = \frac{\lambda_2^2}{180}, \quad f_{s,B} = \min(f_s, 0.99), \qquad (7)$$

where $\lambda_2$ is the secondary dendrite arm spacing, and $K_0$ a constant that depends on the morphology of the mushy region. The last identity is considered to prevent division by zero. In the standard BKC model, the last term in Eq. (2), $S_r$, is omitted compared to the Darcian damping force, $S_u$, [38], and $\boldsymbol{u}_s$ is assumed to be equal to the casting speed (zero in our case).

(b) The hybrid model

Different flow zones exist during the solidification of a metallic alloy: The bulk liquid where $f_s = 0$, the slurry zone where $0 < f_s < f_s^{cr}$, the porous region where $f_s^{cr} < f_s < 1$, and the solid zone where $f_s = 1$, where $f_s^{cr}$ is the value of solid fraction at the interface between slurry and porous regions. In the slurry region ($0 < f_s < f_s^{cr}$), no relative motion is assumed between solid and liquid phases ($\boldsymbol{u} = \boldsymbol{u}_l = \boldsymbol{u}_s$). In the porous region, a rigid dendritic skeleton is formed and the solid velocity will be equal to zero (equal to the given casting speed in case of continuous casting). Furthermore, the viscosity value is considered to be equal to liquid viscosity and Darcy flow becomes significant. The BKC equation only provides accurate results for regions with a relatively large solid fraction, i.e., $f_s > f_s^{cr}$ [40-43]. To account for the transition from slurry to porous, a hybrid model was used by Chang and Stefanescu [13]. In this model, a modified permeability is considered as

$$K^{-1} = (1 - F_\mu)K_0^{-1}\left[\frac{f_{s,B}^2}{(1 - f_{s,B})^3}\right]; \quad K_0 = \frac{\lambda_2^2}{180}, f_{s,B} = \min(f_s, 0.99), \tag{8}$$

where $F_\mu$ is a switching function expressed by [43]

$$F_\mu = 0.5 - \frac{1}{\pi}\arctan[100(f_s - f_s^{cr})], \tag{9}$$

Backerud, et al. [44] recommended the value of 0.27 for $f_s^{cr}$. The variation of the liquid viscosity is also accounted for through a relative viscosity equation [13],

$$\mu_l = \mu_l^0\left\{1 - \frac{f_s F_\mu}{0.3}\right\}^{-2}. \tag{10}$$

Based on the hybrid model assumptions, the relative advection takes the following form [13]:

$$\boldsymbol{S}_r = \begin{cases} 0 & ; f_s < f_s^{cr} \\ \boldsymbol{\nabla} \cdot \left[\rho \frac{f_{s,B}}{(1 - f_{s,B})}\boldsymbol{uu}\right] & ; \text{otherwise} \end{cases}. \tag{11}$$

We found that this discontinues relation leads to errors in the model prediction. Therefore, to consider a smooth transition of $\boldsymbol{S}_r$ between the two regions, we adopted similar blending approach used for the permeability in Eq. (8) for $\boldsymbol{S}_r$ and modified Eq. (11) to

$$S_r = \nabla \cdot \left[(1 - F_\mu)\rho \frac{f_{s,B}}{(1 - f_{s,B})} \boldsymbol{uu}\right]. \tag{12}$$

2.2. *Heat transfer*

To model heat transfer, the energy equation can be described as [38]:

$$\frac{\partial(\rho c_p T)}{\partial t} + \nabla \cdot (\rho \boldsymbol{u} c_p T) = \nabla \cdot (\kappa_{\text{eff}} \nabla T) + \frac{\partial}{\partial t}(\rho L_f f_s) + \nabla \cdot (\rho L_f f_s \boldsymbol{u}) - \nabla \cdot [\rho L_f f_s (\boldsymbol{u} - \boldsymbol{u}_s)], \tag{13}$$

where $L_f$ is the latent heat of solidification, and $c_p$ is the mixture specific heat capacity and can be expressed by

$$c_p = f_s c_{ps} + (1 - f_s) c_{pl}. \tag{14}$$

$\kappa_{\text{eff}}$ is the effective thermal conductivity which is calculated by (assuming a laminar flow)

$$\kappa_{\text{eff}} = g_s \kappa_s + (1 - g_s) \kappa_l. \tag{15}$$

Using Eq. (4), the second term on the right-hand side of Eq. (13) can be recast as

$$\frac{\partial}{\partial t}(\rho L_f f_s) = L_f \rho_s \frac{\partial g_s}{\partial t} = L_f \rho_s \frac{\partial g_s}{\partial T}\frac{\partial T}{\partial t}, \tag{16}$$

$$\frac{\partial g_s}{\partial T} = \frac{\rho^2}{\rho_s \rho_l} \frac{\partial f_s}{\partial T}. \tag{17}$$

and $\partial f_s / \partial T$ is calculated analytically based on the chosen microsegregation model for $f_s$ described in section 2.3. For the standard BKC model, $\boldsymbol{u}_s = 0$ in the last term of Eq. (13), and for the modified hybrid model, this term is treated as

$$\nabla \cdot [\rho L_f f_s (\boldsymbol{u} - \boldsymbol{u}_s)] = \nabla \cdot [(1 - F_\mu)\rho L_f f_s \boldsymbol{u}]. \tag{18}$$

2.3. *Mass transfer*

The species transport equation is based on the classic mixture theory proposed by Bennon and Incropera [7],

$$\frac{\partial(\rho C)}{\partial t} + \nabla \cdot (\rho \boldsymbol{u} C) = \nabla \cdot (\rho D \nabla C) + \nabla \cdot \left[\rho f_l D_{l,\text{eff}} \nabla (C_l - C)\right] + \nabla \cdot \left[\rho f_s D_s \nabla (C_l - C)\right]$$
$$- \nabla \cdot [\rho(\boldsymbol{u} - \boldsymbol{u}_s)(C_l - C)], \tag{19}$$

where $D_{l,\text{eff}} = D_l$ assuming laminar flow, and $D$ is the mass diffusivity of solute in the mixture which is given by

$$D = f_s D_s + (1 - f_s) D_l. \tag{20}$$

To close the equations, the following relations are required:

$$C = f_s C_s + (1 - f_s) C_l. \tag{21}$$

Note also that assuming linear liquidus and solidus lines on the phase diagram, the liquidus, $T_{\text{liq}}$, and solidus, $T_{\text{sol}}$, temperatures are given by

$$T_{\text{liq}} = T_{\text{melt}} + m_{\text{liq}} C, \tag{22}$$

$$T_{\text{sol}} = T_{\text{melt}} + m_{\text{liq}} C / K_p, \tag{23}$$

where $T_{\text{melt}}$ is the fusion (or melting) temperature of the pure solvent, $m_{\text{liq}}$ is the slope of the liquidus line in the alloy phase diagram, $K_p$ is the partition coefficient defined by

$$K_p = C_s^* / C_l^*, \tag{24}$$

and the superscript $*$ is used to indicate the value of a property at the solid-liquid interface. Assuming thermodynamic equilibrium at the interface as well as local thermal equilibrium ($T = T^*$),

$$T = T^* = T_{\text{melt}} + m_{\text{liq}} C_l^*. \tag{25}$$

In summary, Eqs. (21)-(25) are a set of 5 equations and 7 unknowns ($f_s$, $C_s$, $C_l$, $C_s^*$, $C_l^*$, $T_{\text{sol}}$, and $T_{\text{liq}}$). Therefore, a microsegregation model accounting for the relative movement between the solid and liquid phase and solute redistribution at the solidification front is crucial to MS modeling [45]. Here, the Lever rule is used for the solidification of alloys. In this model, thermodynamic equilibrium is assumed in the whole solidification system. This means that solute diffusion in both liquid and solid is assumed infinite [7]. Therefore,

$$C_l = C_l^*, \tag{26}$$

$$C_s = C_s^*, \tag{27}$$

closes the governing equations. Using Eqs. (21)-(27), $f_s$, $C_l$, and $C_s$ can be explicitly computed by

$$f_s = \frac{1}{1-K_p}\left[\frac{T-T_{\text{liq}}}{T-T_{\text{melt}}}\right], \tag{28}$$

$$C_l = C[1 + f_s(K_p - 1)]^{-1}, \tag{29}$$

$$C_s = K_p C_l. \tag{30}$$

Again, for the standard BKC model, $\boldsymbol{u}_s = 0$ in the last term of Eq. (19), and for the modified hybrid model, this relative advection term is given by

$$\nabla \cdot [\rho(\boldsymbol{u} - \boldsymbol{u}_s)(C_l - C)] = \nabla \cdot [(1 - F_\mu)\rho(C_l - C)\boldsymbol{u}]. \tag{31}$$

## 3. Experimental benchmark

Carozzani, et al. [30] experimentally and numerically investigated the solidification of Sn-3%Pb alloy in a rectangular mold, using AFRODITE experimental setup. They used the Cellular Automaton Finite Element (CAFE) computational model. The schematic laboratory model is shown in figure 1a which is 100 $mm$ in length, 60 $mm$ in height, and 10 $mm$ in thickness. Fifty temperature sensors monitor temperature changes over time within the mold. On the right and left-hand side of the mold, there are two heat exchangers made of copper. Both heat exchangers are equipped with K6-type temperature sensors. These sensors are marked with FL1 to FL6 on the left exchanger and FR1 to FR6 on the right exchanger. This test was performed under a vacuum of $1kPa$, thus, air movement is not a factor in heat transfer. The radiant heat loss is compensated by the Kirchhoff box; therefore, the condition of thermal insulation can be assumed for the large side plates and the top surface of the mold. The bottom plate is insulated by an aerogel insulator.

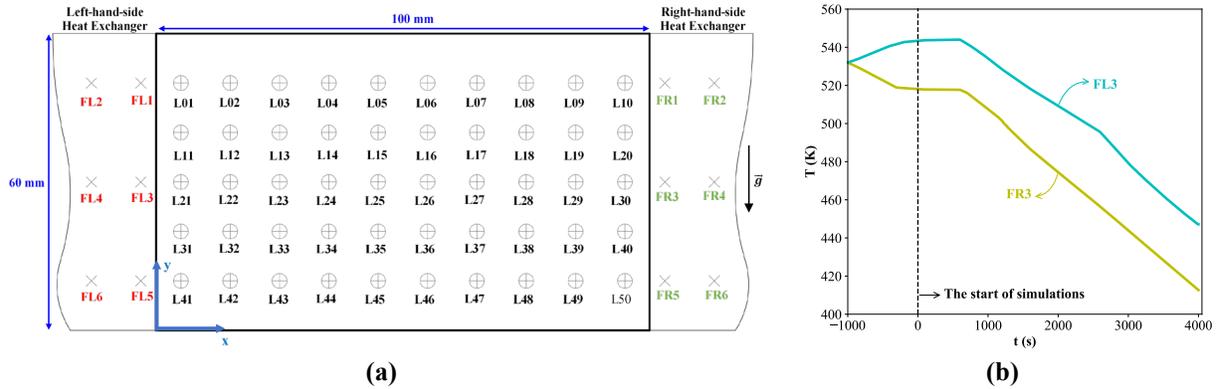

(a)                                 (b)

**Figure 1** a) The schematic set-up of the AFRODITE benchmark, b) The thermal boundary conditions at the left (FL3) and right (FR3) walls [30] versus the simulation time.

A sample of Sn-3%Pb alloy is placed in the mold. It is first heated to a temperature of 533.15 $K$ and held at this temperature for 600 $s$. Under the crucible is an electromagnetic stirrer, used to stir the melt. The aim is to melt the alloy and homogenize the melt. After some time, the stirring is stopped and the time is reset to zero. Then, the left heat exchanger is heated to a temperature of 553.15 $K$, while the right one is cooled to a temperature of 513.15 $K$. These temperatures remain constant for 1000 $s$. Due to this temperature difference, a flow is created by natural convection. After 1000 $s$, the right and left heat exchangers are cooled at a constant rate of -0.03 $Ks^{-1}$ until solidification ends. Thus, there is a temperature difference of 40 K between the right and the left side during the entire solidification (figure 1b). To save the computation time, the whole 1600 s homogenization process applied in the experiment is not necessary since uniform initial thermophysical properties are chosen as the initial condition in the simulations. However, still a period of simulation before the temperature drops at the left and the right wall is necessary to account for the effect of natural convection. We found that an initial 600 s of simulation is sufficient to establish the steady state natural convection. Therefore, our simulations start 1000 s later than the experiment (figure 1b). The thermos-physical properties of Sn-3%Pb are given in table 1.

## 4. Numerical method

The governing equations with the initial and boundary conditions are solved using the open-source software OpenFOAM dev 6 (www.openfoam.org) based on the Finite Volume Method (FVM). The code development has been performed starting with the "solidificationFoam" solver [46] and applying considerable modifications and amendments. The pressure-velocity coupling is treated by the Pressure-Implicit with Splitting of Operators (PISO) method [47] including 2 pressure correction and 5 solidification loops. The detailed flowchart of the solver is reported in Appendix A.

**Table 1** Sn-3%Pb thermo-physical properties [30].

| Parameters | Symbols | Values |
|---|---|---|
| Solid thermal conductivity | $\kappa_s$ | $55\ Wm^{-1}K^{-1}$ |
| Liquid thermal conductivity | $\kappa_l$ | $33\ Wm^{-1}K^{-1}$ |
| Solid specific heat capacity | $c_{p,s}$ | $209\ Jkg^{-1}K^{-1}$ |
| Liquid specific heat capacity | $c_{p,l}$ | $243\ Jkg^{-1}K^{-1}$ |
| Density | $\rho_l, \rho_s$ | $7130\ kgm^{-3}$ |
| Latent heat | $L_f$ | $56140\ Jkg^{-1}$ |
| Mass diffusivity in the Liquid | $D_l$ | $3.5\times10^{-9}\ m^2s^{-1}$ |
| Mass diffusivity in the Solid | $D_s$ | $0\ m^2s^{-1}$ |
| Dynamic viscosity | $\mu_l$ | $2\times10^{-3}\ kgm^{-1}s^{-1}$ |
| Solute expansion coefficient | $\beta_C$ | $-5.3\times10^{-1}$ |
| Thermal expansion coefficient | $\beta_T$ | $9.5\times10^{-5}\ K^{-1}$ |
| Secondary arm spacing | $\lambda_2$ | $9\times10^{-5}\ m$ |
| Initial temperature | $T_0$ | $531.75\ K$ |
| Fusion temperature | $T_{melt}$ | $505.15\ K$ |
| Liquidus line slope | $m_{liq}$ | $-128.6089\ K$ |
| Partition coefficient | $K_p$ | $0.0656$ |

The time derivatives are discretized by the first-order implicit "Euler" scheme [47] and the gradient terms using the second-order "Gauss linear" [47]. In the "Gauss linear" scheme, the "Gauss" keyword specifies the standard finite volume discretization of the Gaussian integration which requires the interpolation of values from cell centers to face centers. This interpolation is done by the second-order "linear" scheme, here. For the discretization of the advection terms, the bounded "Gauss linear upwind" scheme, which employs upwind interpolation weights with an explicit correction based on the local cell gradient and is second-order accurate as shown by Warming and Beam [48], is used. The "Gauss-linear" scheme is adopted for the diffusion terms. The systems of discretized equations are solved by the Geometric Agglomerated Algebraic Multi-Grid (GAMG) solver with the GaussSeidel smoother for the pressure and the Stabilized Preconditioned Bi-Conjugate Gradient (PBiCGStab) algorithm with the Diagonal Incomplete LU decomposition (DILU) pre-conditioning for the other variables. The convergence criterion at each time step is set as the normalized residual tolerance of $10^{-8}$ for the energy, species transport, and momentum equations, and $10^{-6}$ for the pressure.

For the numerical simulation, a uniform grid is generated within the rectangular mold with a cell size of 1 $mm$ × 1 $mm$ based on the grid-independence study, reported in Appendix A. The no-slip condition was set as the velocity boundary condition for all four walls. The thermal insulation condition has been considered for the top and bottom walls, and the temperature at the left- and right-hand side walls is set according to figure 1b. The melt is initially at rest with the nominal composition and temperature of $T_0$.

## 5. Results and discussion

Due to the complexity of MS phenomena and the large number of governing equations involved, a systematic procedure is required to validate a new solver, avoiding the error-hiding effect of different sub-models. For this purpose, the solver can first be verified under limiting conditions. This is performed in section 5.1 against an analytical solution for alloy solidification. Then, in section 5.2, the capabilities and limitations of the present modeling approaches in predicting different features of MS maps are scrutinized using a standard experimental benchmark.

### 5.1. Verification against the 1D alloy-solidification analytical solution

For pure-material solidification with a sharp interface between the melt and solidified regions, analytical solutions have been obtained for the 1D Stefan problem using different levels of simplifications, see e.g., Refs. [21, 49]. In the case of alloy solidification consisting of three different regions, namely melt, mushy, and solid zones, the situation is more complicated. For a 1D semi-infinite domain, Cho and Sunderland [19] derived an analytical solution assuming that the solid mass fraction in the mushy zone is a linear function of the distance (the spatial coordinate). However, microsegregation models usually assume the mass fraction to be a function of the temperature rather than distance, see section 2. Ozisik and Uzzell [22] extended this solution to the linear microsegregation rule in terms of temperature. Nevertheless, this solution was derived for an ideal heat sink with zero diameter in a cylindrical geometry, assuming a constant heat generation rate. These features impose uncertainties when the computational domain is constructed for the validation of a numerical solver. Therefore, we extend the analytical solution of Cho and Sunderland [19] to the linear microsegregation law for a 1D planar semi-infinite domain and a constant chilled-wall temperature condition in Appendix B.

To verify our numerical solver against this analytical solution, the Sn-3%Pb melt with the initial temperature of $T_0 = 905\ K$ and composition of $C_0 = 0.03\ (C_0 < C_\alpha)$ within a semi-infinite domain, which is chosen as a wide rectangle of size $2\ mm \times 2\ m$ discretized by a uniform $2 \times 2000$ grid, is subjected to a chilled boundary with a constant temperature of $T_w = 300\ K$ at the left side, where the solidification starts. The Neumann zero-gradient condition is imposed at the other boundaries. For the general numerical model of section 2, physical properties within the mushy region vary by the change in $f_s$, e.g., by Eq. (14); However, for the analytical solution, these properties are assumed constant. Therefore, for the verification in this section, the properties of both liquid and solid states should be the same which leads to constant properties within the mushy zone, and the numerical model becomes compatible with the analytical model assumptions. The physical properties of the Sn-3%Pb alloy are adopted from table 1 with the difference that the physical properties of the solid are taken to be equal to those of the liquid. The other simplified assumptions of the analytical solution, applied to the general model for this case, are outlined in Appendix B. Based on Eqs. (22) and (23), the solidus and liquidus temperatures used for the analytical solution are $T_{\text{sol}} = 469.04\ K$ and $T_{\text{liq}} = 501.19\ K$.

The comparison of the present numerical simulation using the Linear microsegregation law given by Eq. (B1) with the present analytical solution, detailed in Appendix B, is reported in figure 2. The positions of the solid and liquid front versus time are illustrated in figure 2b. In the numerical simulation, these fronts are taken as $f_s = 0.95$ and $f_s = 0.05$ iso-surfaces, respectively. The analytical solution is given by Eqs. (B15) and (B16) of Appendix B, where their parameters are calculated by Eqs. (B26), (B28), and (B29). There is an excellent agreement between the present analytical solution and numerical simulations. For the sake of comparison, the numerical solution using the Lever rule described by Eq. (28) is also plotted in figure 2. The Lever rule is a non-linear function of temperature and generally results in a higher solid mass fraction prediction than the linear rule within the mushy zone. Owing to the simplifying assumptions and the uniformity of all thermo-physical properties for this case, the energy equation and temperature distribution do not depend on the microsegregation model. As a result, the temperature distributions predicted by both models coincide with each other, as shown in figure 2a, and closely agree with the analytical solution described by Eqs. (B25)-(B27). This comparison manifests that the numerical result perfectly matches the analytical solution which shows the validity of the numerical solver at the limiting condition.

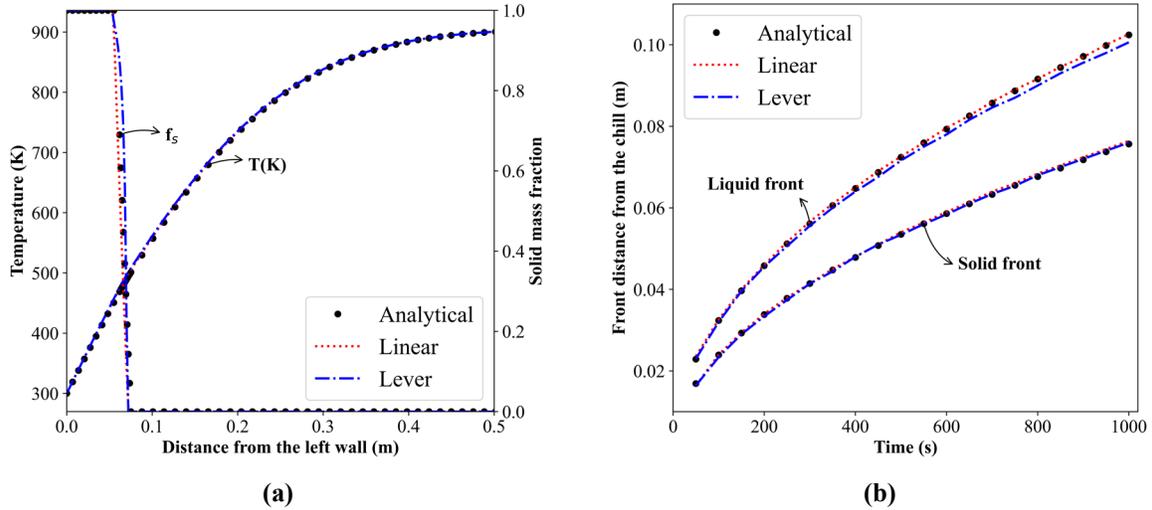

**Figure 2** The verification: The comparison of the present numerical simulations using two microsegregation models, the linear and lever rules, against the analytical solution. a) Temperature and solid mass fraction distribution along the domain at $t = 500\ s$, and b) the solid and liquid front positions, $S_s(t)$ and $S_l(t)$, versus time.

## 5.2. *Macrosegregation in the AFRODITE benchmark*

To validate the MS solver for a real solidification problem, including all fluid flow, heat, and mass transfer phenomena, the present numerical predictions are assessed against experimental measurements for the AFRODITE benchmark. Figure 3 compares the numerical results using the two permeability models, i.e., the standard BKC and hybrid, and experimental data [30], where the temperature and solid mass fraction variation over time at different probe locations in the mold (see also figure 1) are reported. According to figure 3a, after the chilling commences (at about $t = 1200\ s$), the trend of temperature variation at each probe changes distinctly due to the complex physics of the problem. Both models capture these complicated temperature records in reasonable agreement with the experiment. Of course, some phenomena like recalescence, characterized by about 1 K temperature rise at about $t = 2520\ s$ in the inset of the figure 3a, are not captured by the present numerical simulations. The prediction of recalescence necessitates multiscale models [30] due to the effect of grain microstructure change on macroscopic parameters. This is beyond the scope of the present work. In addition, the difference between temperature records predicted by the standard and hybrid models is negligible. The deviation between the models in terms of the solid mass fraction is larger; however, it is still limited to less than 2%.

Figure 4 shows the final MS map. The comparison is made between the standard BKC and hybrid models against the experimental X-ray imaging. The Segregation Degree (SD) is calculated by: $[(C - C_0)/C_0] \times 100\%$. According to this figure, it is established that both models capture the channel segregates at the lower-right part of the ingot (the region bounded with the white dashed box) in accordance with the experimental observation. However, the number and direction of channels predicted by the hybrid model are in much closer agreement with the experiment, compared to the ones predicted by the standard model. In addition, only the hybrid model could predict the major positive segregation zones (bounded by the black dashed boxes) which are evident in the experimental image. The reasons behind the improved MS map prediction by the hybrid model and other MS map characteristics are analyzed carefully in the rest of this section.

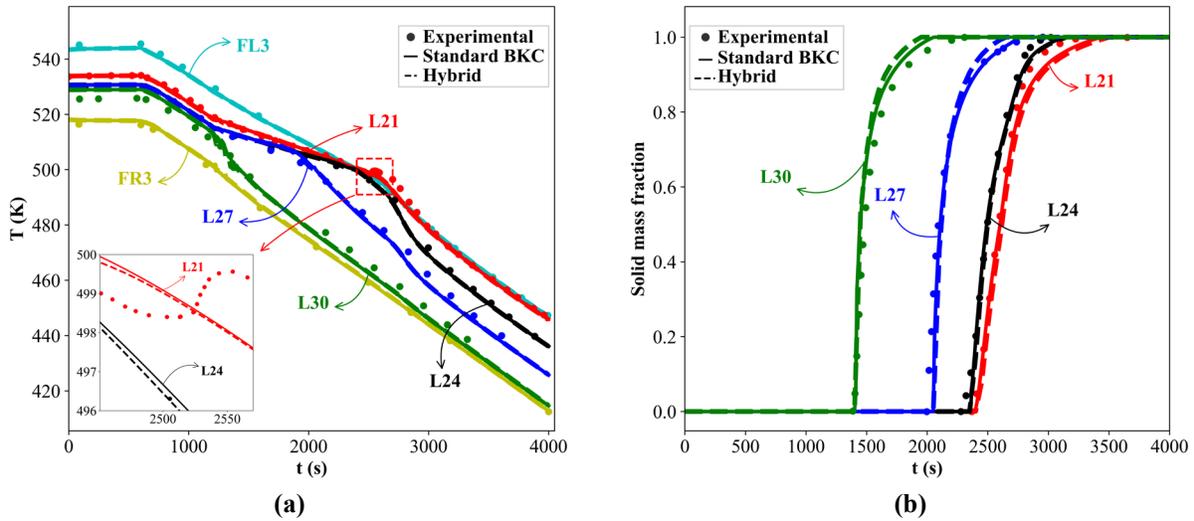

**Figure 3** The validation: The present numerical results using the standard BKC and hybrid models against the experimental data (Sn-3%Pb) [30]: a) Temperature and b) solid mass fraction variation over time at different probe locations, introduced in figure 1.

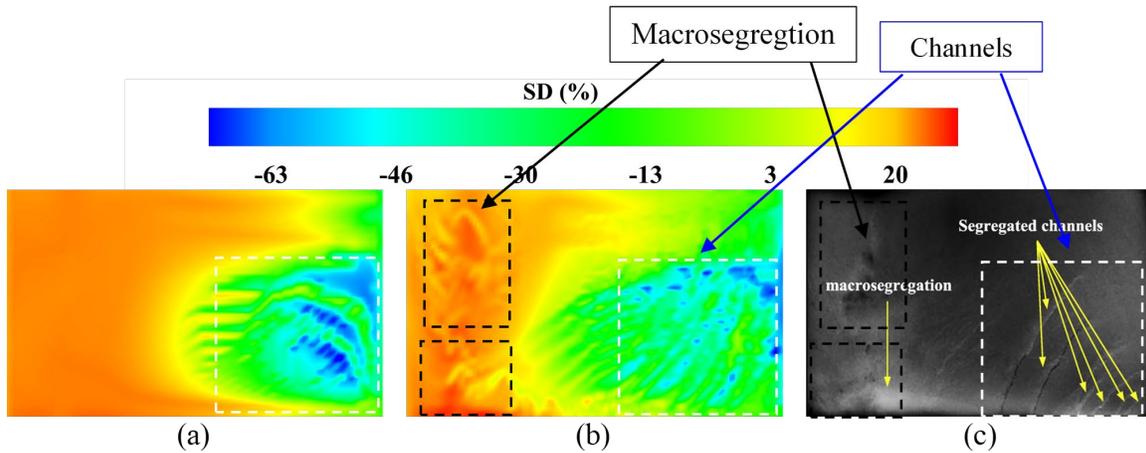

(a)                          (b)                          (c)

**Figure 4** The final MS map: a) Standard BKC model, b) hybrid model, and c) experimental X-ray analysis [32]. The colors in parts a and b show the segregation degree contours. The light and dark regions in the X-ray image (c) are indicative of lead-rich and lead-lean areas, respectively. The main channel segregation region is indicated by the white dashed box and major lead-rich (positive segregation) areas are identified by black dashed boxes.

For a more quantitative analysis, the MS diagrams, i.e., the final SD profiles at different mold heights, are illustrated in figure 5. According to this figure, at the right part of the ingot, see the right part of the profiles at $y = 25\ mm$ and $y = 35\ mm$, where the solidification starts, the amplitude of the SD profile oscillations is larger for the standard model. This shows that the standard model predicts deeper segregation channels than the hybrid one. The aforementioned region corresponds to the zone bounded by the white dashed box in figure 4. Nevertheless, the frequency of oscillations, i.e., the number of channels, is greater for the hybrid model. The other main feature that can be seen in figure 5, is the larger positive SD predicted by the hybrid model near the left wall of the ingot, with a local maximum located at about $x = 0.02\ m$.

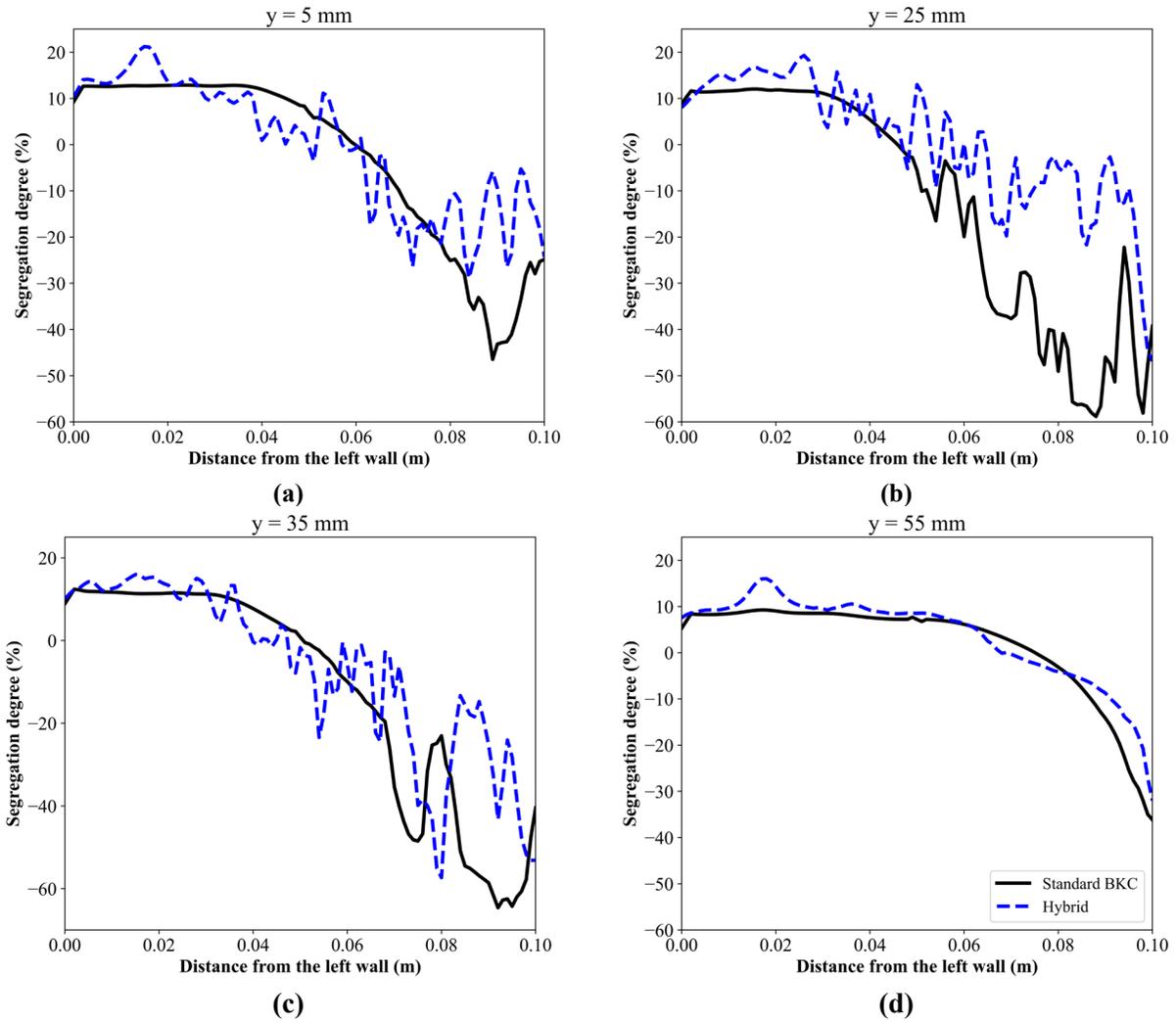

**Figure 5** The final SD profiles at different mold heights using the standard BKC and hybrid models: (a) $y = 5\ mm$, (b) $y = 25\ mm$, (c) $y = 35\ mm$, and (d) $y = 55\ mm$.

To analyze the trend of SD variations, the factors contributing to the SD variation should be considered. During solidification, the solute is pushed ahead of the solidification front. Therefore, the concentration generally decreases behind the solidification front and increases ahead of it. In general, three phenomena have major roles in the level of SD predicted by an MS solver:

1) Back-diffusion which depends on the diffusion in the solid and is dominant at the beginning of the solidification (near the chill source), where the velocity due to the natural convection in the mush is still low. A higher back-diffusion to solid leads to a higher SD.

2) Advection which is more dominant in the middle of the solidification after the diffusion dominant area. Stronger advection washes away the solute-rich region ahead of the solidification front and promotes diffusion from the solid, lowering the SD towards negative values.

3) Concentration pile-up which is simply the conservation of solute mass and is dominant at the final stages of the solidification, i.e., if SD declines at the middle, it should rise at the end.

From the modeling point of view, the back-diffusion strength is governed by the microsegregation law. Here, the same microsegregation model, i.e., the Lever rule, is used for all simulations and imposes the same effect. Therefore, the deviation between the standard and hybrid models in figure 5 is attributed to the advection and pile-up factors. To analyze these factors, additional information on the fluid dynamics of the solidification process is provided in figure 6 and figure 7. In figure 6, the solid mass fraction contours with superimposed flow streamlines, showing the flow direction and vortices, at four different times have been illustrated. In figure 7, the profiles of the solid mass fraction and velocity magnitude at the central line of the mold ($y = 30\ mm$) are plotted at a time instance near the middle of solidification when advection effect is dominant among the aforementioned factors. According to figure 6, a strong clockwise recirculating flow structure is formed due to the thermo-solutal buoyancy force. It is worth noting that the liquid density of the melt varies as solidification proceeds. This is because Pb has a larger density than Sn. Thus, during the solidification, when the solute (Pb) is being pushed ahead of the solidification front, the alloy melt density increases. In other words, the melt with a higher Pb concentration is heavier and moves to the bottom of the mold in the mushy region and its adjacent melt region. The cooling effect of the left wall also decreases the temperature and increases the density of the adjacent fluid. This further augments the downward flow on the solidification front.

The aligned thermal and solutal convections produce a strong clockwise recirculating flow in the mold.

As introduced earlier, in the middle of the solidification, the advection effect on SD is dominant. According to figure 6, when the solidification front is located near the center of the mold, e.g., at $t = 2250\ s$, the velocity magnitude over the solidification front ($f_s < 0.05$) predicted by the hybrid model is greater than the one by the standard model. This is because the hybrid model accounts for the slurry zone near the mush front ($0 < f_s < f_s^{cr}$) and the melt flow penetrates more into the mushy zone, resulting in a stronger advection at the solidification front, compared to the standard model. This stronger advection predicted by the hybrid model causes a lower SD at central mold regions ($0.03 < x < 0.05\ m$), especially at section $y = 5\ mm$ and $y = 35\ mm$ in figure 7, consequently, the subsequent pile-up effect leads to a higher SD on the left side of the mold, compared to the predictions by the standard model. The maximum SD occurring at about $x = 0.02\ m$ in figure 5 or figure 7 is also in accordance with figure 6 which shows that the solidification ends somewhere near $x = 0.02\ m$ (see figure 6, $t = 2500\ s$) and the pile-up effect is pretty large.

On the other hand, this is more intricate to explain the phenomena occurring in the initial stages of solidification. For instance, based on figure 7, near the right wall of the mold, the standard model results in a larger SD and deeper channel segregates while the hybrid model predicts a greater number of channels. These features cannot be analyzed based on the effect of thermo-solutal advection strength parallel to the solidification front.

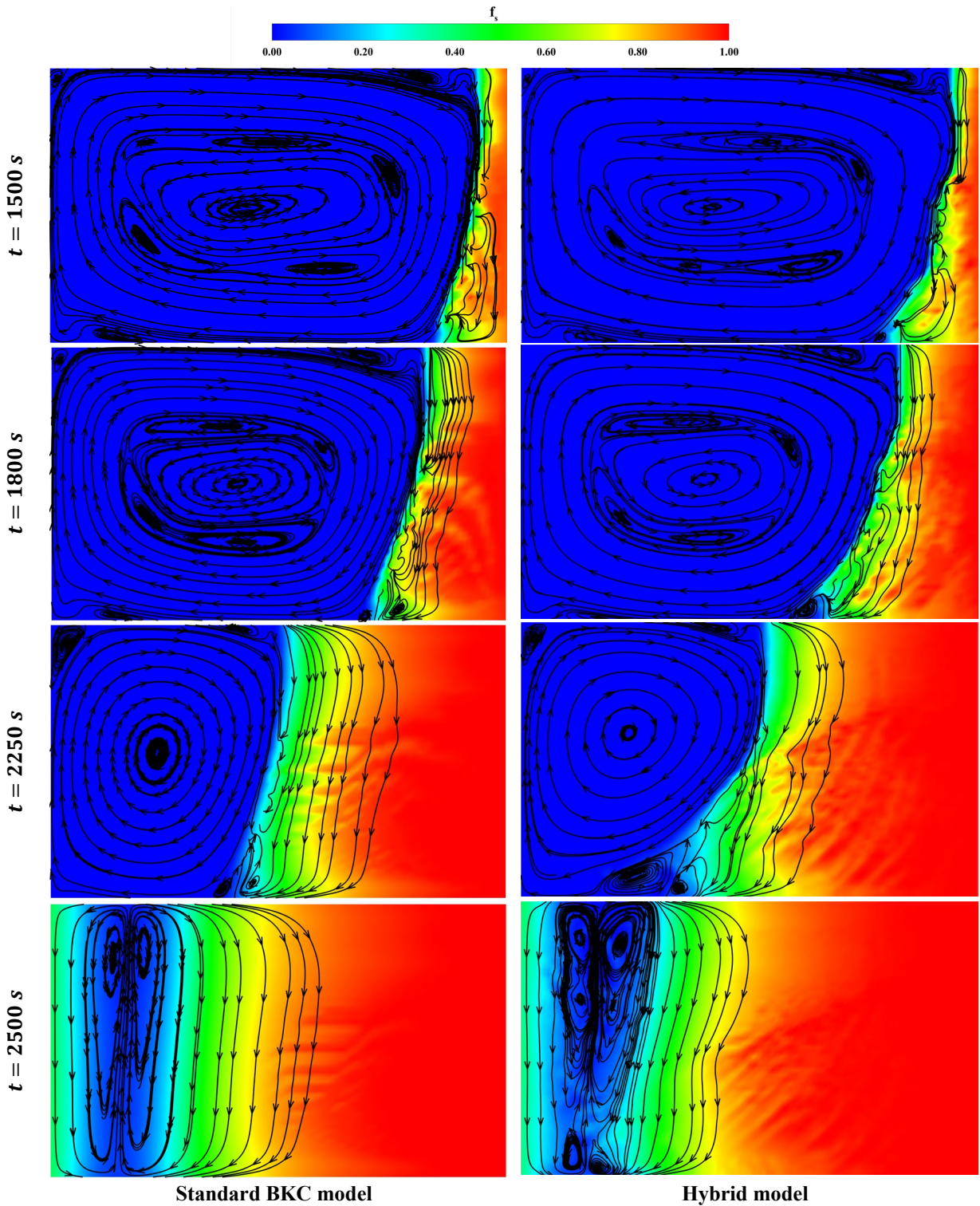

**Figure 6** The solid mass fraction contours with superimposed flow streamlines at different instances: The comparison of the standard BKC (left) and hybrid (right) models.

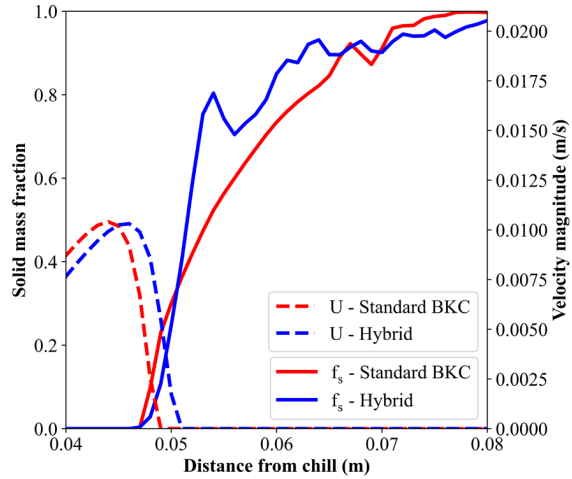

**Figure 7** The solid mass fraction and velocity magnitude profiles at at the mold center ($y = 30\ mm$) and $t = 2250\ s$.

Justifying the segregation channels observed for the standard and hybrid models in figure 5 involves examining miscellaneous contributing factors. The regions with a high potential for channel formation within columnar structures are established due to the interplay of several driving forces [4, 50]. One of the factors contributing to the formation of the potential zones for channel segregates is the re-melting process [50]. There are two main situations in which re-melting occurs: 1) Local reheating due to the hot bulk fluid transported by convection, and 2) concentration increase in the liquid surrounding the dendrites due to the flow of solute-rich liquid into these regions which increases the local liquidus temperature. The present case is especially prone to the second re-melting mechanism owing to Sn-3%Pb alloy's steep liquidus slope; Even minor concentration gradients can cause a considerable drop in the liquidus temperature. In either case, the re-melting can be identified by the local negative sign of $\partial f_s/\partial t$, which is provided by the numerical solution. In order to check whether re-melting happens in the present case, the regions of negative $\partial f_s/\partial t$ predicted by both models are compared in figure 8. According to this figure, at initial stages of solidification when the mushy region traverses the right part of the mold, there are many more re-melting spots in the hybrid model results. This is the main factor underlying the greater number of channels predicted by the hybrid model.

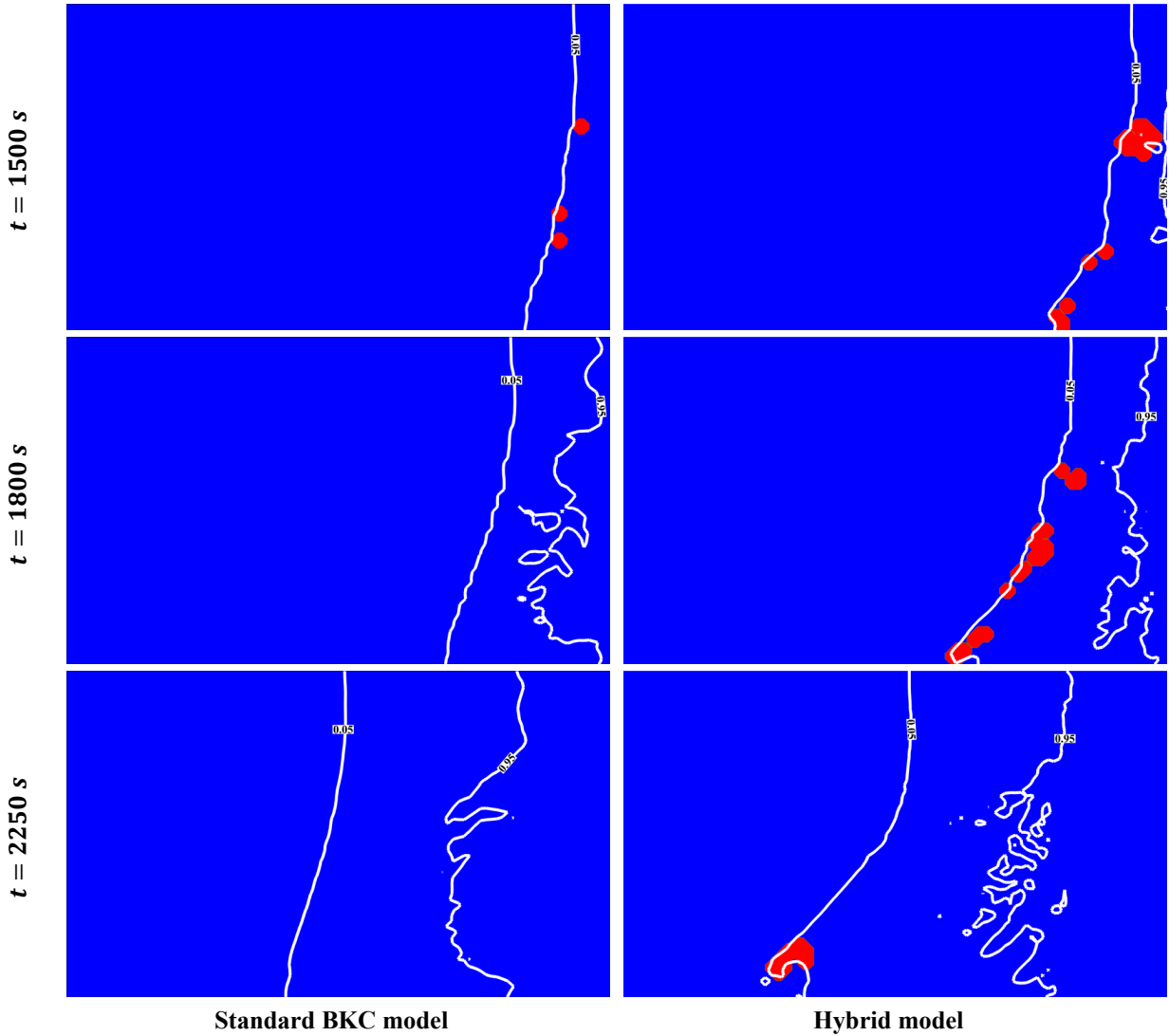

**Figure 8** The re-melting regions ($\partial f_s / \partial t < 0$) shown by red color and $f_s = 0.05$ and $f_s = 0.95$ iso-surfaces, indicating the mushy zone borders, at different time instances: The comparison of the standard BKC (left) and hybrid (right) models.

The other primary source is linked to instabilities in the solidification front caused by bulk convection and movement of interdendritic liquid, which is regulated by permeability within the mushy zone. Several studies tried to connect the presence of the flow instabilities to a critical mushy-zone Rayleigh number [4, 51, 52], which is the ratio of the thermo-solutal buoyancy force to the mush permeability friction force. However, this criterion needs the non-trivial choice of several characteristic parameters and calibration of the critical Rayleigh number for a specific alloy type. Here, we directly visualize the vortical structures and flow instabilities formed in the vicinity of the solidification front, based on the detailed flow-field information obtained from the numerical solution, to identify the regions and instances of channel formation during the solidification process. The vortical flow structures predicted using the standard and hybrid models are compared during the solidification process in figure 6. According to this figure, the flow structures are characterized by a large dominant clockwise recirculating flow generated by the thermo-solutal buoyancy force. Within this recirculating zone and far from the solidification front, several secondary vortices form and dissipate during the solidification. Strong flow instabilities lead to the break-up and coalescence of these vortices throughout the solidification. However, they do not affect the solidification front and have negligible influence on channel formation. If the secondary vortical structures are formed near the mushy region boundaries, they can make small cavities on the solidification front. These cavities have the potential to convert into segregate channels since the outflow from the preamble mushy region can short-circuit through these areas, rather than passing through the neighboring porous medium of a higher resistance to the flow. Likewise, the inflow to the mushy region rises through these cavities due to the capillary effect. These intensified flow rates cause the cavities to grow to channels aligned with the local flow direction. According to figure 6, the hybrid model results are accompanied with slightly more strong instabilities in terms of small vortices and levels of flow perturbations near the solidification front, e.g., see the vortex formed near the bottom of the mold on the solidification front in figure 6. As a result, the difference in the flow instabilities is the next factor triggering the different MS pattern predicted by both models. However, this factor is deemed secondary compared to re-melting since, based on our experience, the formation of strong secondary vortices near the solidification front is more frequent when the thermal and solutal buoyancy forces are not in the same direction, where the local imbalance between these forces can lead to stronger flow instabilities near the front. This is not the case in the present problem, and the aligned thermal and solutal buoyancy generates a stable flow over the mushy region.

The third factor introduced by Mehrabian, et al. [53] is the flow-solidification interaction term, defined as $\boldsymbol{u}_l \cdot \boldsymbol{\nabla} C_l$, which accounts for the interdendritic flow-solidification interaction [53, 54]. This term serves to delineate two distinct solidification regions: A suppressed solidification region, characterized by a negative value of this term, indicating that interdendritic flow is in the opposite direction to the liquid concentration gradient, and an accelerated solidification region, where this term takes a positive value, indicating that interdendritic flow is in the same direction to the liquid concentration gradient. The contours of the flow-solidification interaction term are reported in figure 9. Channels are only likely to form in the suppressed solidification region ($\boldsymbol{u}_l \cdot \boldsymbol{\nabla} C_l < 0$), where a perturbation-induced increase in local flow is reinforced by the resulting suppressed solidification, leading to continued channel growth. Conversely, channels do not grow in the accelerated solidification region ($\boldsymbol{u}_l \cdot \boldsymbol{\nabla} C_l > 0$), where an increase in local flow intensity due to a perturbation is counteracted by accelerated solidification. As can be observed in figure 9, when the solidification front ($f_s = 0.05$) passes across the right half of the mold, the results by the standard model show a number of spots near the solidification front with very large negative values of $\boldsymbol{u}_l \cdot \boldsymbol{\nabla} C_l$. These spots appear and sustain exactly where the channel segregates are observed on the MS map (figure 4). These spots are also present in the predictions by the hybrid model; Nevertheless, due to the stronger convection predicted in the slurry region by the hybrid model and the reduced concentration gradient in flow direction over the front, the magnitude of $\boldsymbol{u}_l \cdot \boldsymbol{\nabla} C_l$ within these spots is much smaller compared to those by the standard model. Therefore, the larger negative flow-solidification interaction term is responsible for the deeper channels predicted by the standard model.

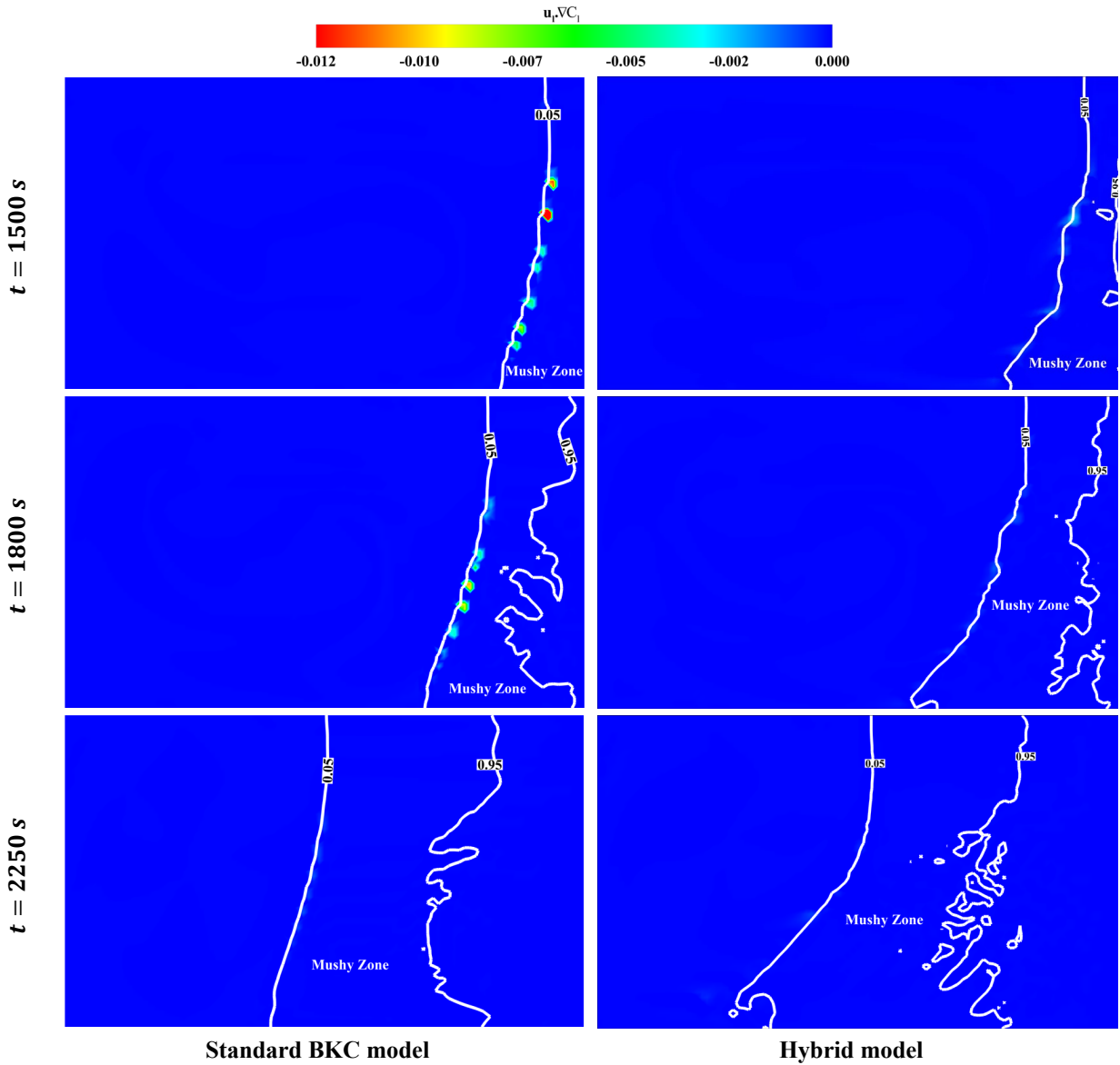

**Figure 9** The flow-solidification interaction term ($\mathbf{u}_l \cdot \nabla C_l$) and $f_s = 0.05$ and $f_s = 0.95$ iso-surfaces, indicating the mushy zone borders, at different instances: The comparison of the standard BKC (left) and hybrid (right) models. Only the negative contour level values are shown for the sake of clarity.

## 6. Conclusion

In the present research, a systematic benchmarking of MS solvers was sought to avoid the error-hiding effect of different sub-models. For this purpose, an analytical solution for the alloy solidification Stefan problem was obtained, assuming the linear microsegregation law. Then, this solution was adopted to verify an OpenFOAM MS solver capability to predict alloy solidification correctly in the limiting condition of the absence of fluid flow and mass transfer. After that, the MS solver was used to assess the performance of the standard Blake-Kozeny-Carman and a modified hybrid permeability laws for the Sn-3%Pb test case of the state-of-the-art AFRODITE solidification benchmark. The evaluation included the ability to predict temperature and composition records, as well as the important features of the MS map such as the channel segregation and regions of distinct positive and negative SD. We slightly modified the relative advection terms in the hybrid model to incorporate a continuous variation of these terms from the slurry to porous regions. In contrast to the standard model, the hybrid model captures the peak SD due to the pile-up effect at the end of the solidification in accordance with the experimental X-ray image. This improvement was attributed to the stronger enriched melt advection and wash-off along the solidification front owing to the correct incorporation of the slurry to porous behavior transition with the hybrid model. On the other hand, both models could predict the channel segregation. However, the channels predicted by the standard model were deeper than those by the hybrid closure while those by the hybrid model are greater in number. To find the reason behind this observation, different factors contributing to channel formation, including flow instabilities, re-melting, and flow-solidification interaction, were considered and analyzed. It was found that the re-melting is the main cause of the greater number of channels predicted by the hybrid model while the flow-solidification interaction justifies the deeper channels seen in the standard model predictions. The coincidence of the location of the large negative flow-solidification interaction parameter and the final channel segregates corroborated this idea.

**Declaration of Interests:** The authors report no conflict of interest

**Supplementary Material**

The supplementary material provides videos for the animated comparison of the solidification process using the standard and hybrid models.

**Acknowledgments**

This work was partially supported by Mobarakeh Steel Company [grant number 48565554].

**Appendix A: The grid-independence check**

The grid resolution of 1 $mm$ is chosen based on the recommendation by Carozzani, et al. [30] for this test case. To further check this mesh resolution, the results of predictions using 3 different grid resolutions are reported in figure 10. This figure proves the independency of the temperature records from the chosen computational grid size.

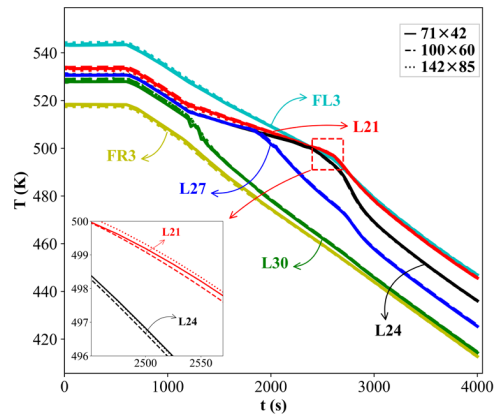

**Figure 10** The grid-independence test: The numerical prediction of temperature record at several probe locations using different grid resolutions.

**Appendix B: Analytical solution for the 1D alloy-solidification Stefan problem**

An analytical solution for the alloy solidification considering three different regions, including liquid, mush, and solid, is obtained assuming the 1D Stefan problem shown in figure 11. The governing equations are derived by applying a number of simplifications to the general form of the energy equation, Eq. (13). These simplifying assumptions are:

1) Advection is neglected.

2) Thermo-physical properties within the melt, mushy, and solid regions are assumed to be constant. However, they can be different for different regions, except for density (the density difference induces considerable advection).

3) The solute concentration distribution is uniform ($D_l$ and $D_s$ are infinite), and $T_{liq}$ and $T_{sol}$ are constant.

4) The microsegregation model follows a linear rule within $0 \leq f_s \leq 1$ as

$$f_s = f_{su}\left[1 - \frac{T - T_{End}}{T_{liq} - T_{End}}\right]; \quad T_{End} = \begin{cases} T_{sol} & ; C < C_\alpha \\ T_{eut} & ; C_\alpha \leq C < C_{eut} \end{cases}. \tag{B1}$$

For a hypoeutectic alloy ($C_\alpha < C < C_{eut}$), $f_{su}$ is a constant depending on the alloy composition and outside the eutectic range ($C < C_\alpha$), $f_{su}$ is unity.

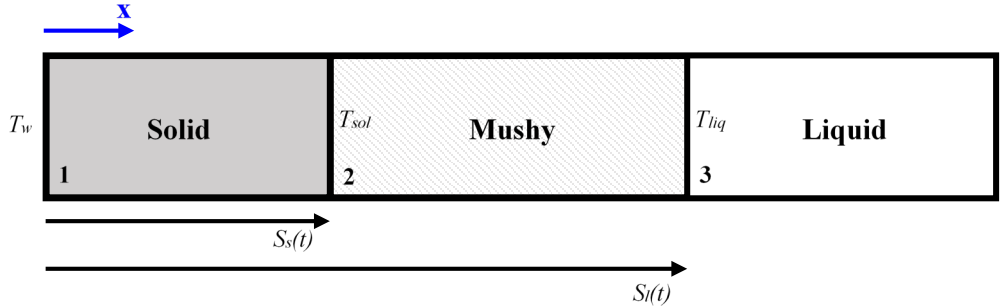

**Figure 11** The schematic figure of the 1D alloy solidification Stefan problem. $S_s$ and $S_l$ denote the position of the solid and liquid interfaces.

Using these simplifications, the energy equation for all three regions shown in figure 11 can be written as

$$\frac{\partial^2 T_1}{\partial x^2} = \frac{1}{\alpha_1}\frac{\partial T_1}{\partial t} \quad ; 0 \leq x < S_s(t), \tag{B2}$$

$$\frac{\partial^2 T_2}{\partial x^2} + \frac{\rho L_f}{\kappa_2}\frac{\partial f_s}{\partial t} = \frac{1}{\alpha_2}\frac{\partial T_2}{\partial t} \quad ; S_s(t) \leq x < S_l(t), \tag{B3}$$

$$\frac{\partial^2 T_3}{\partial x^2} = \frac{1}{\alpha_3}\frac{\partial T_3}{\partial t} \quad ; x \geq S_l(t), \tag{B4}$$

where $\alpha$ is the thermal diffusivity defined by $\alpha = k/\rho c$. The initial conditions are

$$T_1(x,0) = T_2(x,0) = T_3(x,0) = T_0, \tag{B5}$$

$$S_s(0) = S_l(0) = 0, \tag{B6}$$

and the boundary conditions are

$$T_1(0,t) = T_w, \tag{B7}$$

$$T_1(S_s,t) = T_2(S_s,t) = T_{sol}, \tag{B8}$$

$$T_2(S_l,t) = T_3(S_l,t) = T_{liq}, \tag{B9}$$

$$T_3(\infty,t) = T_0. \tag{B10}$$

Since $S_s(t)$ and $S_l(t)$ are unknown, two additional equations are required to close the system. For this purpose, the energy balances at the solid-mushy interface, $x = S_s(t)$, and mushy-melt interface, $x = S_l(t)$, are presented as

$$-\kappa_1 \frac{\partial T_1}{\partial x} + \kappa_2 \frac{\partial T_2}{\partial x} + \rho L_f \frac{dS_s}{dt}(1 - f_{su}) = 0 \ ; x = S_s(t), \tag{B11}$$

$$-\kappa_2 \frac{\partial T_2}{\partial x} = -\kappa_3 \frac{\partial T_3}{\partial x} \ ; x = S_l(t). \tag{B12}$$

The non-dimensional variables can be defined by

$$\theta = \frac{T - T_w}{T_0 - T_w}; \tau = \frac{\alpha_1 t}{L^2}; y = \frac{x}{L}; \psi_s = \frac{S_s}{L}; \psi_l = \frac{S_l}{L}, \tag{B13}$$

where $L$ is an arbitrary length scale. A similarity solution can be obtained for the governing equations by assuming the similarity variable as

$$\xi = \frac{y}{2\sqrt{\tau}}, \tag{B14}$$

and the mushy zone non-dimensional boundary locations as

$$\psi_s(\tau) = 2\lambda_s\sqrt{\tau}, \tag{B15}$$

$$\psi_l(\tau) = 2\lambda_l\sqrt{\tau}, \tag{B16}$$

where $\lambda_l$ and $\lambda_s$ are the constants that have to be determined. Using Eqs. (B13)-(B16), the governing equations, Eqs. (B2)-(B12), can be recast as

$$\frac{d^2\theta_1}{d\xi^2} = -2\xi \frac{d\theta_1}{d\xi}, \tag{B17}$$

$$\frac{d^2\theta_2}{d\xi^2} = -2\xi \left(\frac{1+\theta_c}{\alpha_{12}}\right)\frac{d\theta_2}{d\xi}, \quad \theta_c = \frac{L_f f_{\text{su}}}{c_{p,2}(T_{\text{liq}} - T_{\text{End}})}, \tag{B18}$$

$$\frac{d^2\theta_3}{d\xi^2} = -\frac{2\xi}{\alpha_{13}} \frac{d\theta_3}{d\xi}, \tag{B19}$$

$$\theta_1(0) = 0, \tag{B20}$$

$$\theta_1(\lambda_s) = \theta_2(\lambda_s) = \theta_{\text{sol}}, \tag{B21}$$

$$\theta_2(\lambda_l) = \theta_3(\lambda_l) = \theta_{\text{liq}}, \tag{B22}$$

$$\frac{d\theta_1}{d\xi} = \kappa_{12}\frac{d\theta_2}{d\xi} + \frac{2\lambda_s L_f(1-f_{\text{su}})}{c_{p,1}(T_0 - T_w)} \quad ; \xi = \lambda_s, \tag{B23}$$

$$\frac{d\theta_2}{d\xi} = \kappa_{23}\frac{d\theta_3}{d\xi} \quad ; \xi = \lambda_l, \tag{B24}$$

where $\alpha_{ij} = \alpha_j/\alpha_i$ and $\kappa_{ij} = \kappa_j/\kappa_i$.

By direct integration of Eqs. (B17)-(B19), satisfying the initial and boundary conditions, Eqs. (B20)-(B22), the temperature distribution in each region is obtained as

$$\theta_1 = \frac{\theta_s}{\text{erf}(\lambda_s)}\text{erf}(\xi), \tag{B25}$$

$$\theta_2 = \frac{(\theta_{liq} - \theta_{\text{sol}})\text{erf}(m\xi) + \theta_{\text{sol}}\text{erf}(m\lambda_l) - \theta_{\text{liq}}\text{erf}(m\lambda_s)}{\text{erf}(m\lambda_l) - \text{erf}(m\lambda_s)}, m = \sqrt{\frac{1+\theta_c}{\alpha_{12}}}, \tag{B26}$$

$$\theta_3 = \frac{(\theta_{\text{liq}} - 1)}{\text{erfc}\left(\frac{\lambda_l}{\sqrt{\alpha_{13}}}\right)}\text{erfc}\left(\frac{\xi}{\sqrt{\alpha_{13}}}\right) + 1. \tag{B27}$$

Then, substituting Eqs. (B25)-(B27) into Eqs. (B23) and (B24) yields

$$\theta_{\text{sol}}\frac{e^{-\lambda_s^2}}{\text{erf}(\lambda_s)} - m\kappa_{12}(\theta_{\text{liq}} - \theta_{\text{sol}})\left[\frac{e^{-(m\lambda_s)^2}}{\text{erf}(m\lambda_l) - \text{erf}(m\lambda_s)}\right] - \frac{(1-f_{\text{su}})L_f\sqrt{\pi}}{c_{p,1}(T_0 - T_w)}\lambda_s = 0, \quad (B28)$$

$$m(\theta_{\text{liq}} - \theta_{\text{sol}})\left[\frac{e^{-(m\lambda_l)^2}}{\text{erf}(m\lambda_l) - \text{erf}(m\lambda_s)}\right] + \kappa_{23}\frac{(\theta_{\text{liq}} - 1)}{\text{erfc}\left(\frac{\lambda_l}{\sqrt{\alpha_{13}}}\right)}\frac{e^{-\left(\frac{\lambda_l^2}{\alpha_{13}}\right)}}{\sqrt{\alpha_{13}}} = 0. \quad (B29)$$

Eqs. (B28) and (B29) are solved in a coupled manner to determine the parameters $\lambda_l$ and $\lambda_s$. Here, to solve these nonlinear coupled equations, a hybrid iterative algorithm is employed, leveraging a combination of MINPACK's "hybrd" and "hybrj" algorithms [55] using "scipy.optimize.fsolve()" in python. Knowing $\lambda_l$ and $\lambda_s$, the temperature distribution is presented by Eqs. (B25)-(B27) and Eqs. (B15) and (B16) give the position of the mushy region borders over time.